\documentclass[10pt,twocolumn]{article}

\usepackage[margin=0.9in]{geometry}
\usepackage{graphicx}
\usepackage[T1]{fontenc}
\usepackage[utf8]{inputenc}
\usepackage[hidelinks, colorlinks=true, linkcolor = black, citecolor = black, urlcolor = blue]{hyperref}
\usepackage{titling}
\usepackage{authblk}
\usepackage[]{abstract}
\usepackage[font=small,labelfont=bf]{caption}
\usepackage{titlesec}
\usepackage{multibib}
\newcites{Methods}{References}
\usepackage[float]{}

\titleformat*{\section}{\large\bfseries}
\titleformat*{\subsection}{\normalsize\bfseries}

\usepackage{newunicodechar}

\newunicodechar{×}{$\times$}
\newunicodechar{⁻}{$^{-}$}
\newunicodechar{¹}{$^{1}$}
\newunicodechar{⁶}{$^{6}$}
\newunicodechar{⁷}{$^{7}$}
\newunicodechar{⁸}{$^{8}$}
\newunicodechar{⁹}{$^{9}$}
\newunicodechar{⁽}{$^{(}$}
\newunicodechar{³}{$^{3}$}
\newunicodechar{⁾}{$^{)}$}

\pretitle{\vspace{-15mm}\begin{center}\large\bfseries}
\posttitle{\par\end{center}}

\title{An ultrafast electro-optic light source with sub-cycle precision}

\author[,1]{David R. Carlson\thanks{david.carlson@nist.gov}}
 \author[1]{Daniel D. Hickstein}
 \author[1]{Wei Zhang}
 \author[1,2]{Andrew~J.~Metcalf}
 \author[1]{Franklyn Quinlan}
 \author[1,2]{Scott~A.~Diddams}
 \author[,1,2]{Scott B. Papp\thanks{scott.papp@nist.gov}}
 \affil[1]{Time and Frequency Division, National Institute of Standards and Technology, 325 Broadway, Boulder, CO, 80305}
 \affil[2]{Dept. of Physics, University of Colorado, 2000 Colorado Ave, Boulder, CO, 80309}
\date{\vspace{-3mm}\small\today}

\makeatletter
\newbox\abstract@box
\renewenvironment{abstract}
  {\global\setbox\abstract@box=\vbox\bgroup
     \hsize=\textwidth\linewidth=\textwidth
    \small
    \vspace{-6mm}
    \quotation}
  {\endquotation\egroup}
\expandafter\def\expandafter\@maketitle\expandafter{\@maketitle
  \ifvoid\abstract@box\else\unvbox\abstract@box\if@twocolumn\vskip1.5em\fi\fi}
\makeatother

\makeatletter
\renewcommand{\fnum@figure}{Fig. \thefigure}
\makeatother

\begin{document}

\begin{abstract}
Controlling femtosecond optical pulses with temporal precision better than one cycle of the carrier field has a profound impact on measuring and manipulating interactions between light and matter. We explore pulses that are carved from a continuous-wave laser via electro-optic modulation and realize the regime of sub-cycle optical control without a mode-locked resonator. Our ultrafast source, with a repetition rate of 10 GHz, is derived from an optical-cavity-stabilized laser and a microwave-cavity-stabilized electronic oscillator. Sub-cycle timing jitter of the pulse train is achieved by coherently linking the laser and oscillator through carrier--envelope phase stabilization enabled by a photonic-chip supercontinuum that spans up to 1.9 octaves across the near infrared. Moreover, the techniques we report are relevant for other ultrafast lasers with repetition rates up to 30 GHz and may allow stable few-cycle pulses to be produced by a wider range of sources.
\end{abstract}

\maketitle

Ultrafast lasers produce femtosecond-duration pulses of light and can operate as frequency combs to provide a time and frequency reference spanning the optical and microwave domains~\cite{cundiff_colloquium:_2003}. For applications across science and technology, ultrafast pulse-trains with repetition frequencies of 10~GHz and higher are needed for sampling or exciting high-speed or transient events and making precision measurements across octaves of bandwidth. Yet simultaneously creating broad spectral coverage, low-noise performance, and timing synchronization into the femtosecond domain and below is an unmet challenge.  One approach to producing multi-gigahertz pulse trains is through electro-optic modulation (EOM) of a continuous-wave (CW) laser~\cite{kobayashi_highrepetitionrate_1972, kourogi_wide-span_1993}. These optical pulse generators, henceforth referred to as ``EOM combs'', first gained interest nearly fifty years ago due to their broad tunability, reliability, high power per mode, and spectral flatness~\cite{torres-company_optical_2014, li_electro-optical_2014, millot_frequency-agile_2015, beha_electronic_2017}.  However, they exhibit electronic-oscillator-limited noise that has prevented femtosecond-level timing stabilization, and they have been limited to relatively narrow spectral bandwidths due to the technical difficulties of broadening gigahertz-rate, low-energy pulses.

Here we present broadly applicable techniques for resolving the challenges associated with multi-gigahertz pulse trains originating from EOM combs and other ultrafast sources, such as Kerr microresonators \cite{herr_temporal_2013} and modelocked semiconductor lasers \cite{tilma_recent_2015}. In particular, we generate electro-optic pulse trains at 10 and 30~GHz with approximately 1-ps pulse durations, and show how to spectrally broaden them to octave bandwidths and to temporally compress them to less than three optical cycles (15~fs) in nanophotonic silicon-nitride (Si$_3$N$_4$, henceforth SiN) waveguides. To deliver a stable ultrafast source timed with sub-cycle precision, our work introduces an EOM-comb configuration with optical-cavity stabilization of the CW laser and high-$Q$ microwave-cavity stabilization of a 10-GHz electronic oscillator. The microwave oscillator is phase locked to the CW laser via $f$--$2f$ stabilization of the carrier--envelope offset, enabling complete knowledge of the EOM-comb frequencies to 17~digits. The result is an optical frequency comb of $\sim$28,000 sub-hertz-linewidth modes spanning the near infrared with commensurate sub-cycle timing jitter, and is enabled by key experimental techniques from areas of physics, photonics, and ultralow-noise microwave electronics

To date, this level of spectral coverage and stability has only been achieved with lower-repetition-rate mode-locked lasers.  Now though, EOM combs are poised to broaden the scope of ultrafast sources by providing access to a new regime of extensible construction, high repetition rates, broad bandwidth, low noise, and ultrashort pulse durations. In particular, applications including coherent communications~\cite{marin-palomo_microresonator-based_2017}, astronomical spectrograph calibration~\cite{li_laser_2008, steinmetz_laser_2008}, biological imaging \cite{camp_jr_chemically_2015}, and spectroscopy~\cite{coddington_dual-comb_2016, heinecke_optical_2009, duran_electro-optic_2016} will benefit from these new EOM-based sources.

\begin{figure*}[h]
\centering
\includegraphics[width=\linewidth]{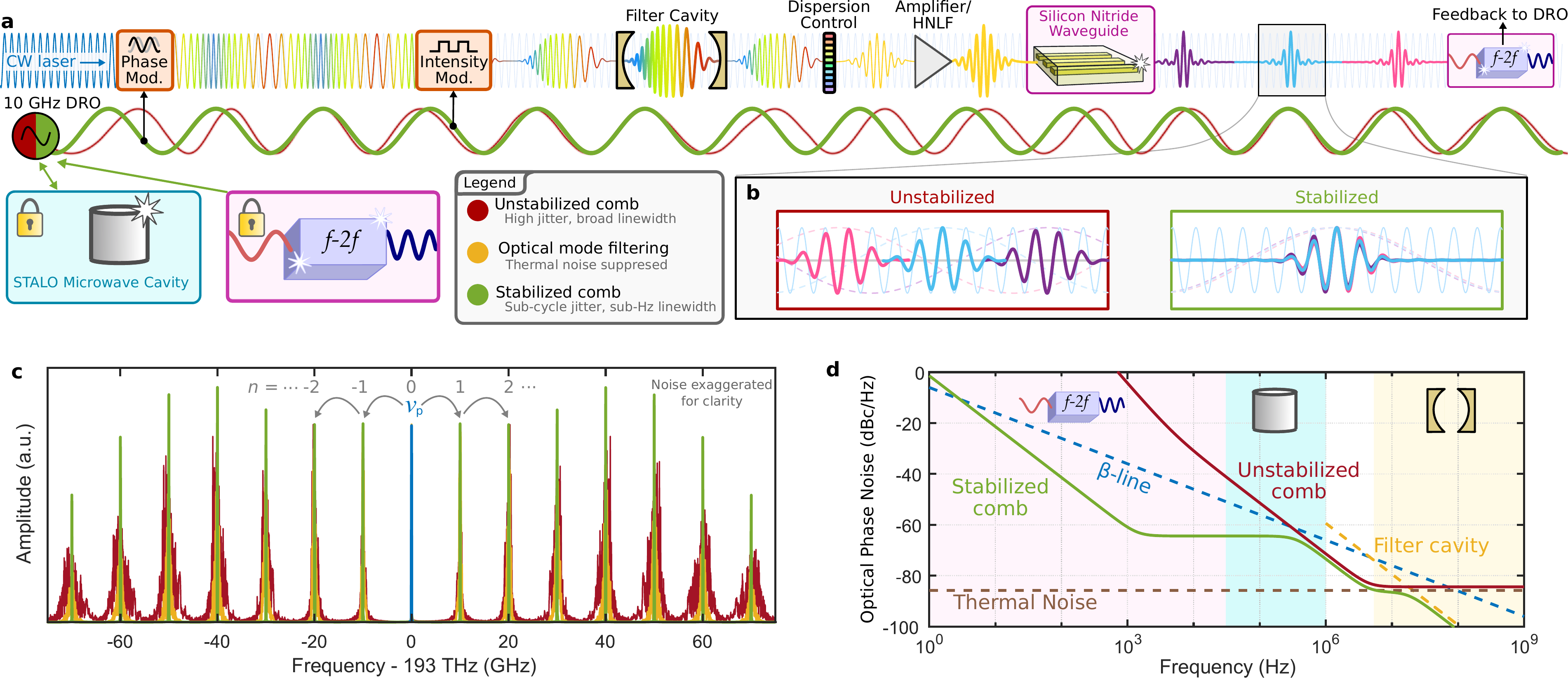}
\caption{Carving ultrafast pulses from a CW laser with sub-cycle precision. a) A chirped pulse train is derived from a \mbox{1550-nm} continuous-wave (CW) laser using electro-optic phase and intensity modulation driven by a 10-GHz dielectric resonant oscillator (DRO) that is locked to a high-Q microwave cavity in the stabilized-local-oscillator (STALO) configuration. The pulse train is then optically filtered by a Fabry--Perot cavity to suppress electronic thermal noise on the comb lines before spectral broadening in highly nonlinear fiber (HNLF) followed by a silicon-nitride waveguide. Octave-spanning spectra allow detection of the comb offset frequency in an $f$--$2f$ interferometer that is used to stabilize the DRO output. b) Without stabilization, the microwave-derived pulse train exhibits large pulse-to-pulse timing jitter relative to the CW carrier.  When the drive frequency is stabilized by feedback from the comb offset frequency and the STALO cavity, sub-optical-cycle phase coherence between successive pulses is achieved. Note: the stabilized pulses are shown with zero carrier--envelope offset, though this is not generally the case. c) Frequency-domain picture of EOM-comb generation.  The unstabilized comb (red) exhibits large noise multiplication as the mode number $n$ expands about zero.  Mode filtering (yellow) suppresses high-frequency thermal noise. The fully-stabilized comb lines (green) appear as $\delta$-functions because the CW-laser stability is transferred across the entire comb bandwidth. d) Optical phase noise picture of the comb, showing the effects of the $f$--2$f$ stabilization, STALO cavity, and filter cavity.}
\label{fig:overview}
\end{figure*}

\section*{Electro-optic modulation combs}
EOM combs are formed by various combinations of intensity and phase modulation of a CW laser \cite{torres-company_optical_2014}. Here we use a microwave source to drive an intensity modulator placed in series with multiple phase modulators to produce a 50\%-duty-cycle pulse train with mostly linear frequency chirp. In the spectral domain, this process results in a deterministic cascade of sidebands with prescribed amplitude and phase that converts the CW laser power into a frequency comb with a mode spacing given by the microwave driving frequency $f_{\rm{eo}}$ (see Fig.~\ref{fig:overview}c).  The frequency of each resulting mode $n$, counted from the CW laser at frequency $\nu_p$, can then be expressed as $\nu_n = \nu_p \pm nf_{\rm{eo}}$.  Equivalently, the modes can be expressed as a function of the classic offset frequency $f_0$ and repetition rate $f_{\rm{rep}}$ parameters as $\nu_n = f_0 + n^\prime f_{\rm{rep}}$, where now the mode number $n^\prime$ is counted from zero frequency and $ f_{\rm{rep}} = f_{\rm{eo}}$. The total number of modes generated by a single phase modulator is determined by the microwave drive power and the modulator's half-wave voltage, though multiple modulators can be added in series with their microwave phases aligned to linearly increase the number of modes. The chirped-pulse output of the comb can be readily compressed to near the Fourier-transform limit using single-mode optical fiber \cite{kobayashi_optical_1988, metcalf_broadly_2015}.

In order for an EOM comb to achieve phase coherence between the CW pump laser and comb modes across the entire spectrum, it is vital to keep the integrated phase noise of each mode below $\pi$ radians. In the temporal domain this corresponds to sub-cycle timing jitter and for EOM combs this requirement becomes more difficult to achieve as the comb bandwidth is increased due to microwave-noise multiplication~\cite{walls_rf_1975}.  In fact, for octave-spanning spectra at a 10~GHz repetition rate, this multiplication factor is roughly 20,000 and corresponds to an 86~dB increase in phase noise. Thus, reaching the $\pi$-radian threshold with an EOM comb requires careful treatment of the noise at all Fourier frequencies.  

As noted in previous work \cite{beha_electronic_2017, kim_cavity-induced_2017}, broadband thermal noise in the electronic components up to the Nyquist frequency can cause the phase-coherence threshold to be exceeded. A Fabry--Pérot cavity can optically filter the broadband thermal noise fundamentally present in electro-optic modulation, resulting in a detectable carrier--envelope-offset frequency \cite{beha_electronic_2017}. However, the cavity linewidth (typically a few megahertz) places a lower bound on the range of frequencies where this suppression is possible, and therefore, it is necessary to investigate the use of low-noise microwave oscillators as well. This is especially important for the Fourier-frequency range between 100 kHz and the filter-cavity linewidth, where high-gain feedback is technically challenging.  However, sufficient feedback to the oscillator \emph{can} be provided below 100~kHz via stabilization of the comb offset frequency $f_0$, as described in the following sections.

\section*{Results}

A schematic of the stabilized EOM comb is shown in Fig.~\ref{fig:overview}a and a comprehensive system diagram appears in Supplementary Fig.~\ref{fig:schematic}. We use a commercial dielectric-resonator oscillator (DRO) with a nominal operating frequency of 10~GHz and 0.1\% tuning range to drive the modulators. Compared to other commercial microwave sources, the DRO offers improved phase-noise performance in the critical Fourier-frequency range between 100~kHz and 10~MHz. The DRO output is then amplified before driving the phase modulators to produce the typical comb spectrum shown in Fig.~\ref{fig:spectrum}a.

After transmission through an optical-filter cavity to suppress thermal noise, the chirped-pulse output of the EOM comb is compressible to durations as short as 600~fs, depending on the initial spectral bandwidth.  Unfortunately, pulse durations greater than roughly 200~fs pose problems for typical supercontinuum broadening in nonlinear media with anomalous dispersion \cite{dudley_supercontinuum_2006, tamura_fundamentals_2000}. However, if the nonlinear material exhibits \emph{normal} dispersion, broadening due to pure self-phase modulation is known to produce lower-noise spectra due to the suppression of modulation instability \cite{tamura_fundamentals_2000, huang_nonlinearly_2008}. Consequently, we employ a two-stage broadening scheme~\cite{cole_octave-spanning_2016} using a normal-dispersion highly nonlinear fiber (HNLF) to achieve initial spectral broadening and pulse compression, followed by an anomalous-dispersion SiN waveguide for broad spectrum generation. The output of the first-stage HNLF is collimated in free-space and compressed to less than 100~fs using a pair of diffraction gratings. After losses are taken into account, pulses with energies up to 150~pJ are delivered to the SiN waveguide. 

\subsection*{Spectral broadening in photonic waveguides}

\begin{figure}
\centering
\includegraphics[width=\linewidth]{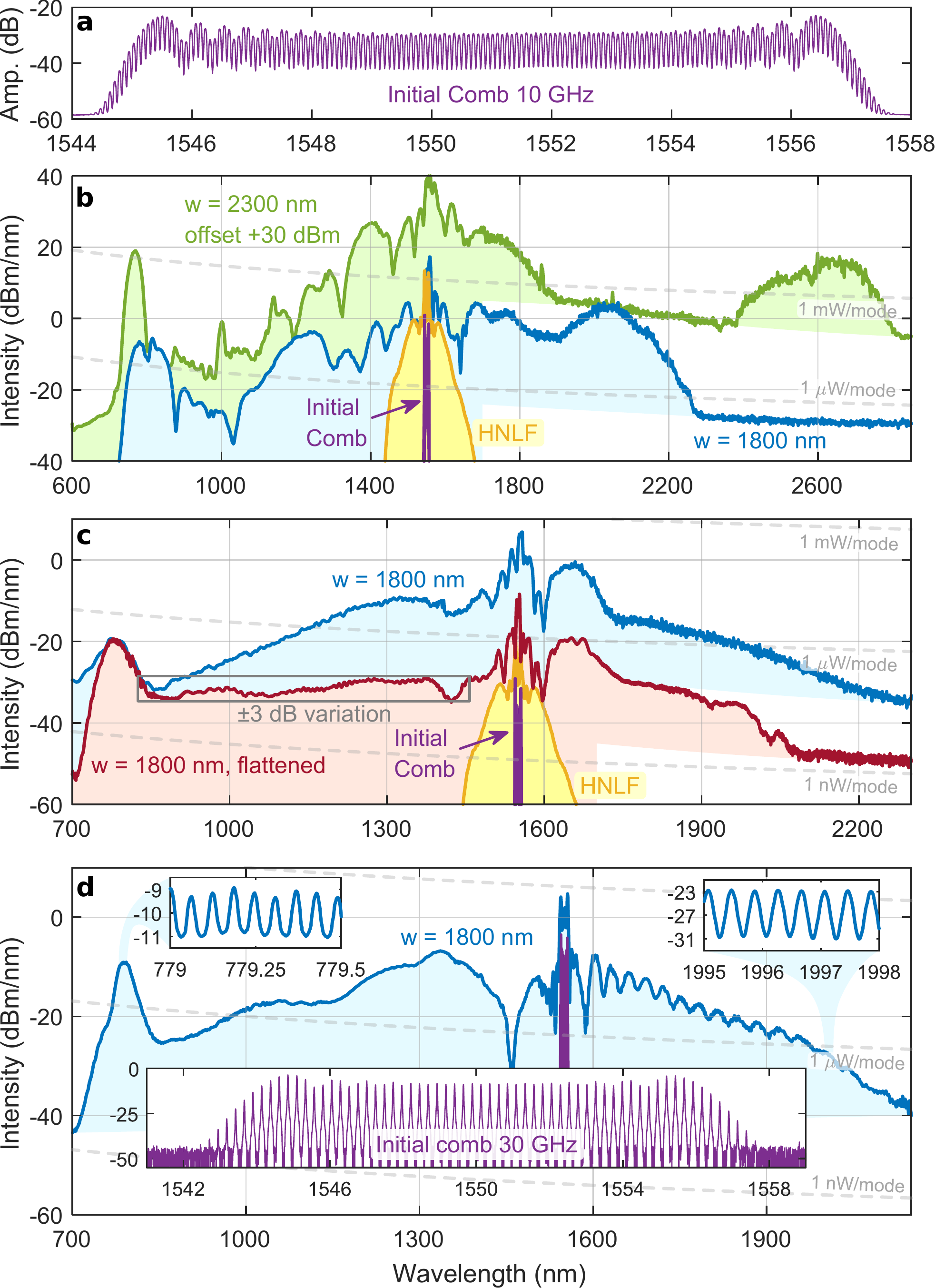}
\caption{High-repetition-rate supercontinuum. a) Spectrum of the 10-GHz EOM comb directly after generation. b) 10-GHz supercontinuum spectra spanning from 750 to 2750~nm for two different silicon-nitride waveguide widths. The spectral intensity is scaled to intra-waveguide levels. Also shown is the spectrum of the first-stage highly nonlinear fiber (HNLF). c) 10-GHz supercontinuum optimized for spectral flatness by reducing incident power. Between 830 and 1450~nm, a flat spectrum ($\pm$3~dB) is produced using a single passive optical attenuator. d) Supercontinuum spectrum from a 30-GHz EOM comb.  Top insets show comb coherence is maintained across the entire spectrum (extinction ratio limited by resolving power of the optical spectrum analyzers used). Bottom inset shows initial spectrum of the 30-GHz EOM comb.  The y-axes in both c) and d) show the spectral intensity obtained in the output fiber.}
\label{fig:spectrum}
\end{figure}

High-repetition-rate lasers ($f_{\rm{rep}}\geq 10$ GHz) produce lower pulse energies for the same average power, making it challenging to use nonlinear broadening to produce the octave bandwidths required for self-referencing. Previous efforts to spectrally broaden such pulses have primarily used nonlinear fibers \cite{millot_frequency-agile_2015, beha_electronic_2017, wu_supercontinuum-based_2013, bartels_10-ghz_2009, kashiwagi_direct_2016} but lithographically patterned waveguides have recently emerged as a serious alternative due to their significantly higher nonlinearity, engineerable dispersion, CMOS-compatible integration, and high-efficiency input coupling \cite{johnson_octave-spanning_2015,zhao_visible--near-infrared_2015, klenner_gigahertz_2016}.  In at least several instances, spectra spanning more than two octaves have been achieved with modest input pulse energies \cite{porcel_two-octave_2017,carlson_self-referenced_2017,hickstein_ultrabroadband_2017}.

Here, input-coupling efficiency to a SiN waveguide of up to 85\% (see Methods) enables a broadband continuum to be generated using pulses from high-repetition-rate ultrafast sources.  Fig.~\ref{fig:spectrum}b shows supercontinuum spectra generated with our 10-GHz EOM comb spanning wavelengths from 750~nm to beyond 2700~nm for two different waveguide geometries. Individual comb lines across the entire bandwidth exhibit a high degree of extinction (50~dB at 1064~nm, see Supplementary Fig.~\ref{fig:cwheterodyne} for data at 775~nm, 1064~nm, and 1319~nm) and do not exhibit any inter-mode artifacts such as sidebands, a common problem when mode filtering is used to convert low-repetition-rate combs to high repetition rates~\cite{braje_astronomical_2008,steinmetz_fabryperot_2009}. 

In order to investigate the scalability to even higher repetition rates, we made additional supercontinuum measurements using a 30-GHz EOM comb, which produced 600-fs, 70-pJ pulses (Fig.~\ref{fig:spectrum}d). Despite the reduction in pulse energy compared to the 10-GHz comb, similar broadband spectra are readily obtained.  In both cases, if the waveguide input power is kept sufficiently low, smooth spectra can be obtained with high power per comb mode.  For applications requiring very flat spectra over broad bandwidths, such as astronomical spectrograph calibration \cite{ycas_demonstration_2012, mccracken_decade_2017}, the supercontinuum light can be easily collected in a single-mode fiber and flattened using a single passive optical attenuator (see Methods).  Under these conditions, fluctuations in spectral intensity can be kept within $\pm$3~dB over wavelengths spanning from 850 to 1450~nm while delivering more than 10~nW per mode in the fiber at 10~GHz.  Improved waveguide-to-fiber output coupling, or free-space collimation combined with an appropriate color filter, could potentially improve the power per comb mode by another factor of 10.

\begin{figure} 
\centering
\includegraphics[width=\linewidth]{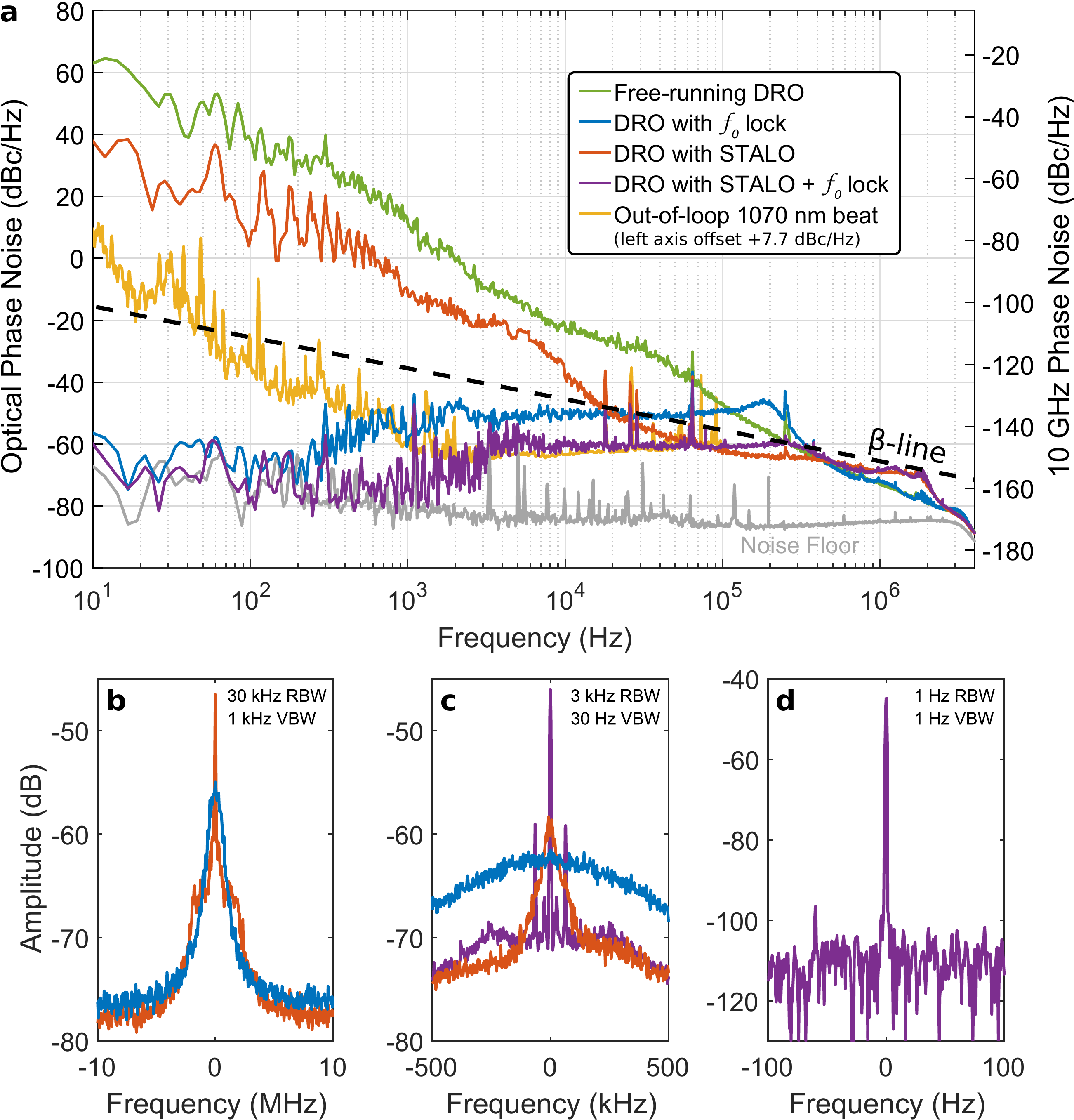}
\caption{EOM-comb phase noise. a) Optical phase noise of the comb offset frequency measured at 775~nm (left axis) and scaled to the 10-GHz repetition rate (right axis) under different locking conditions.  Pre-stabilizing the free-running RF oscillator (DRO) using a high-$Q$ microwave cavity in the stabilized-local-oscillator (STALO) configuration lowers the phase noise by up to 20~dB at frequencies below 500~kHz.  When servo feedback from the optical $f_0$ signal is engaged, a tight phase lock is achieved that suppresses low-frequency noise.  The $\beta$-line indicates the level above which phase noise causes an increase in the comb linewidth.  When both the STALO and $f_0$ locks are engaged, the phase noise remains below the $\beta$-line at all frequencies indicating the coherence of the CW pump laser is faithfully transfered across the entire comb spectrum. b-d) $f_0$ RF beats showing the effects of each feedback loop.  A coherent carrier signal is observed (d) only when both the STALO lock and direct $f_0$ feedback are engaged.}
\label{fig:phasenoise}
\end{figure}

A key advantage of chip-integrated waveguides compared to nonlinear fibers is the ability to readily tune the dispersion properties in order to generate a supercontinuum spectrum optimized for a particular application.  For example, tailored spectra can be obtained in these devices simply by lithographically varying the waveguide width. Here, in order to achieve self-referencing of the EOM comb with an $f$--$2f$ interferometer \cite{jones_carrier-envelope_2000}, the waveguide geometry is chosen to provide supercontinuum light directly at the second harmonic of the 1550-nm pump laser.  This scheme decouples the supercontinuum generation from the frequency doubling of the pump and allows for very stable operation \cite{carlson_self-referenced_2017}. 

The offset frequency is detected with $>$30~dB SNR, suggesting that the scheme of combining normal- and anomalous-dispersion media indeed allows us to overcome some of the difficulties with producing a coherent supercontinuum using pulses longer than a few hundred femtoseconds; see Supplementary Fig.~\ref{fig:offsetvslines} for SNR versus bandwidth. Stabilization of $f_0$ is subsequently accomplished by feeding back to the frequency-tuning port of the DRO.  However, due to optical and electronic phase delay in this configuration, the feedback bandwidth is limited to approximately 200~kHz and thus is insufficient on its own to narrow the comb linewidth set by the multiplied microwave noise of the DRO. 

\subsection*{Microwave noise reduction with STALO cavity}

In order to reach the $\pi$-radian threshold for phase coherence between the CW laser and electronic oscillator, the output of one high-power microwave amplifier is pre-stabilized to an air-filled aluminum microwave cavity in the stabilized-local-oscillator (STALO) configuration \cite{dick_measurement_1989, gupta_high_2004}. By placing the STALO cavity after the amplifier, phase noise introduced by both the DRO and the amplifier is converted to a voltage fluctuation that can be used as a negative-feedback correction signal for the microwave oscillator. Locking the DRO directly to this cavity yields an immediate reduction in phase noise of up to 20~dB at frequencies less than 500~kHz from the carrier (see Fig.~\ref{fig:phasenoise}a).  Increasing the cavity $Q$ or adding an additional low-phase-noise amplifier before the STALO mixer would likely improve the amount of phase-noise reduction that is achievable due to the enhanced sensitivity of the frequency-discrimination signal~\cite{gupta_high_2004}.

In Fig.~\ref{fig:phasenoise}a we use the $\beta$-line~\cite{di_domenico_simple_2010} to distinguish between regimes where the linewidth of the comb offset $f_0$ is adversely affected (phase noise above the $\beta$-line) and where there is no linewidth contribution (phase noise below the $\beta$-line). Having phase noise below the $\beta$-line at all points is approximately equivalent to an integrated phase noise below $\pi$ radians, and thus provides a convenient visual way to assess the impact of noise at different Fourier frequencies. For our EOM comb, the $f_0$ phase noise remains below the $\beta$-line at all frequencies \emph{only} when both the STALO lock and the $f$--$2f$ lock are used in tandem.  Under these conditions, noise arising from the microwave oscillator does not contribute significantly to the comb linewidth and thus, the CW laser stability is faithfully transfered across the entire comb bandwidth.  Equivalently, integrating the phase noise of the fully locked $f_0$ beat (1.17 rad, 10~Hz to 4~MHz) yields a pulse-to-pulse timing jitter of 0.97~fs, implying the microwave envelope coherently tracks the optical carrier signal with sub-cycle precision.

The progression of offset-frequency stabilization is also shown by the beat frequencies in Fig.~\ref{fig:phasenoise}b-d as each lock is turned on.  The coherent carrier seen in the offset frequency when fully stabilized (Fig.~\ref{fig:phasenoise}d) indicates phase coherence has been achieved between individual comb lines across the entire available spectral bandwidth.

\subsection*{Few-cycle pulse generation}
\begin{figure}
\centering
\includegraphics[width=\linewidth]{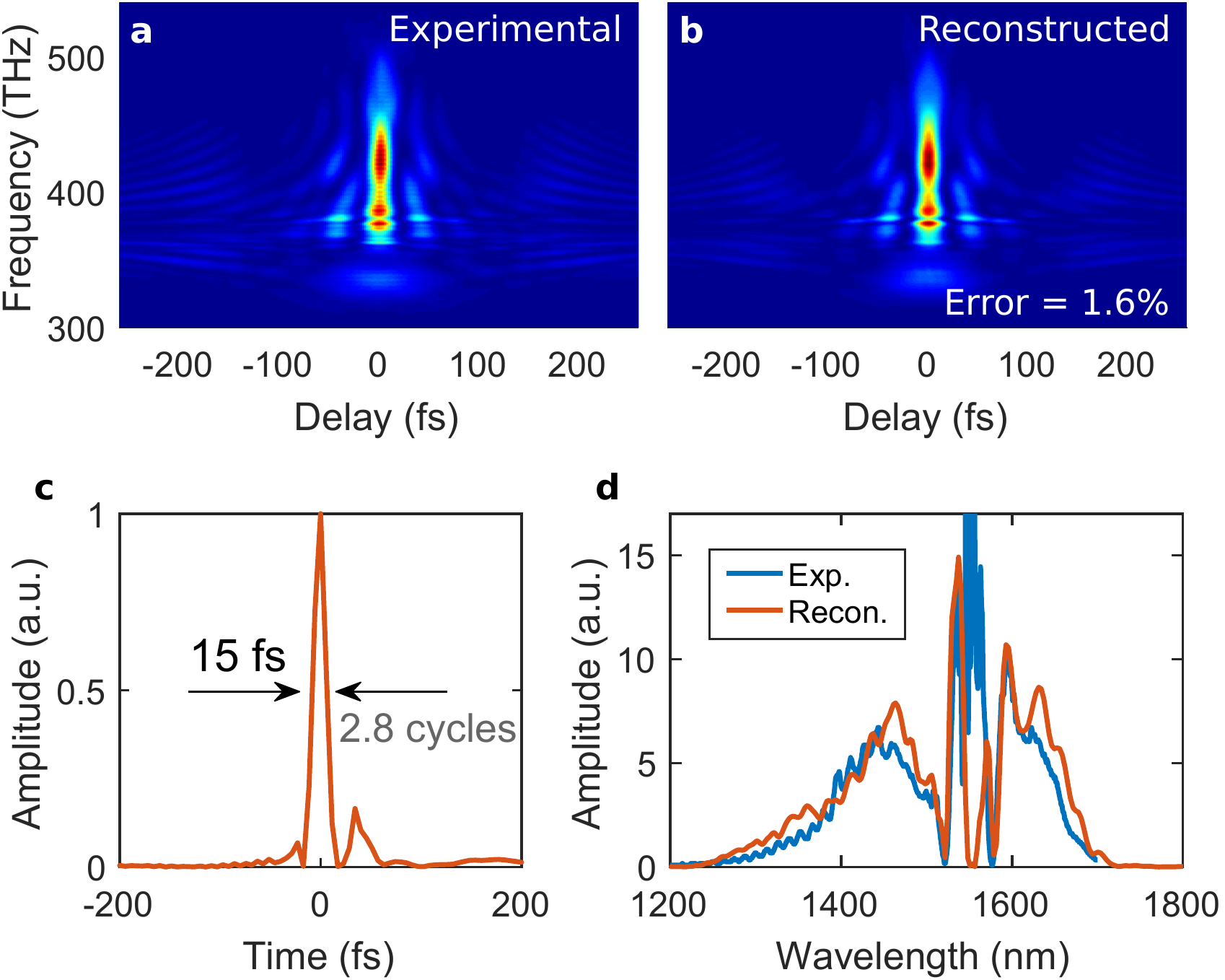}
\caption{Few-cycle pulse generation. a) Experimental and b) reconstructed FROG traces. c) Reconstructed temporal pulse profile with a full-width at half-maximum duration of 15~fs (2.8 optical cycles). c) Comparison of reconstructed and experimental spectra. The quasi-CW spectral wings of the initial comb spectrum near 1550~nm do not contribute significantly to the pulse and thus are not seen in the reconstructed spectrum.  At least 75\% of the total optical power is concentrated in the compressed pulse. More sophisticated amplitude and phase compensation of the initial comb spectrum could allow an even larger fraction of the power to be compressed \cite{jiang_spectral_2007}.}
\label{fig:frog}
\end{figure}

Mode-locked lasers producing few-cycle pulses have had a substantial impact in diverse fields such as chemistry \cite{zewail_femtochemistry:_2000}, attosecond science \cite{corkum_attosecond_2007}, micro-machining \cite{gattass_femtosecond_2008}, and strong-field physics \cite{brabec_intense_2000}.  However, generating stable few-cycle pulses at gigahertz repetition rates with mode-locked lasers is technically challenging, and yet, such pulses are essential for probing the real-time dynamics of nano- and atomic-scale systems where repeatability is impossible or acquisition time is at a premium. For example, coherent Raman imaging of biological samples strongly benefits from transform-limited ultrashort pulses, but the acquisition speed is limited by current megahertz-rate systems \cite{camp_jr_chemically_2015,ideguchi_coherent_2013}. We show here that the use of optical modulators to directly carve a train of $\sim$1~ps pulses from a CW laser provides an effective method for generating few-cycles pulses thanks to the soliton self-compression effect \cite{agrawal_nonlinear_2013, mollenauer_extreme_1983}.  

To demonstrate the ability of EOM combs to produce ultrashort pulses, the pulse power and chirp incident on the SiN waveguide are adjusted such that the launched pulse approaches the threshold peak intensity for soliton fission near the output facet of the chip.  A normal-dispersion single-element aspheric lens is then used to out-couple the light without introducing significant higher-order dispersion. Finally, a 2-cm-long rod of fused silica glass recompresses the pulse to near its transform limit.  

Fig.~\ref{fig:frog} shows the reconstructed pulse profile obtained through frequency-resolved optical gating (FROG)~\cite{trebino_measuring_1997,sidorenko_ptychographic_2016}. Pulse durations of 15~fs (2.8 optical cycles, full width at half maximum) and out-coupled pulse energies in excess of 100~pJ (1~W average power) are readily achievable at a repetition rate of 10~GHz.  Currently, it is not possible to operate the EOM comb with full stabilization while also producing few-cycle pulses as the threshold for soliton fission must be exceeded in order to efficiently generate light for offset-frequency detection.  However, it would be possible, in a future chip design, to utilize an on-chip splitter to both stabilize the offset frequency and generate few-cycle pulses in separate waveguides, as an all-in-one solution to CEO-stabilized pulses at repetition rates in excess of 10~GHz.

\subsection*{Comb performance}
For precision-measurement applications, the absolute accuracy and stability of the comb is an important metric for comb performance. To assess these aspects of our low-noise EOM comb, we compare the 10-GHz repetition rate against the 40$^\mathrm{th}$ harmonic of another independent self-referenced frequency comb operating at a repetition rate of 250~MHz.  By phase locking the appropriate tooth of the reference comb to the same CW laser serving as the pump of the EOM comb, the drift of the CW laser can be completely canceled (see Supplementary Material).  Thus, the measured frequency difference between the two comb repetition rates can be defined solely in terms of the known RF-synthesizer set points and used to assess the absolute frequency-synthesis accuracy of the comb without any data fitting or drift correction.

\begin{figure}
\centering
\includegraphics[width=\linewidth]{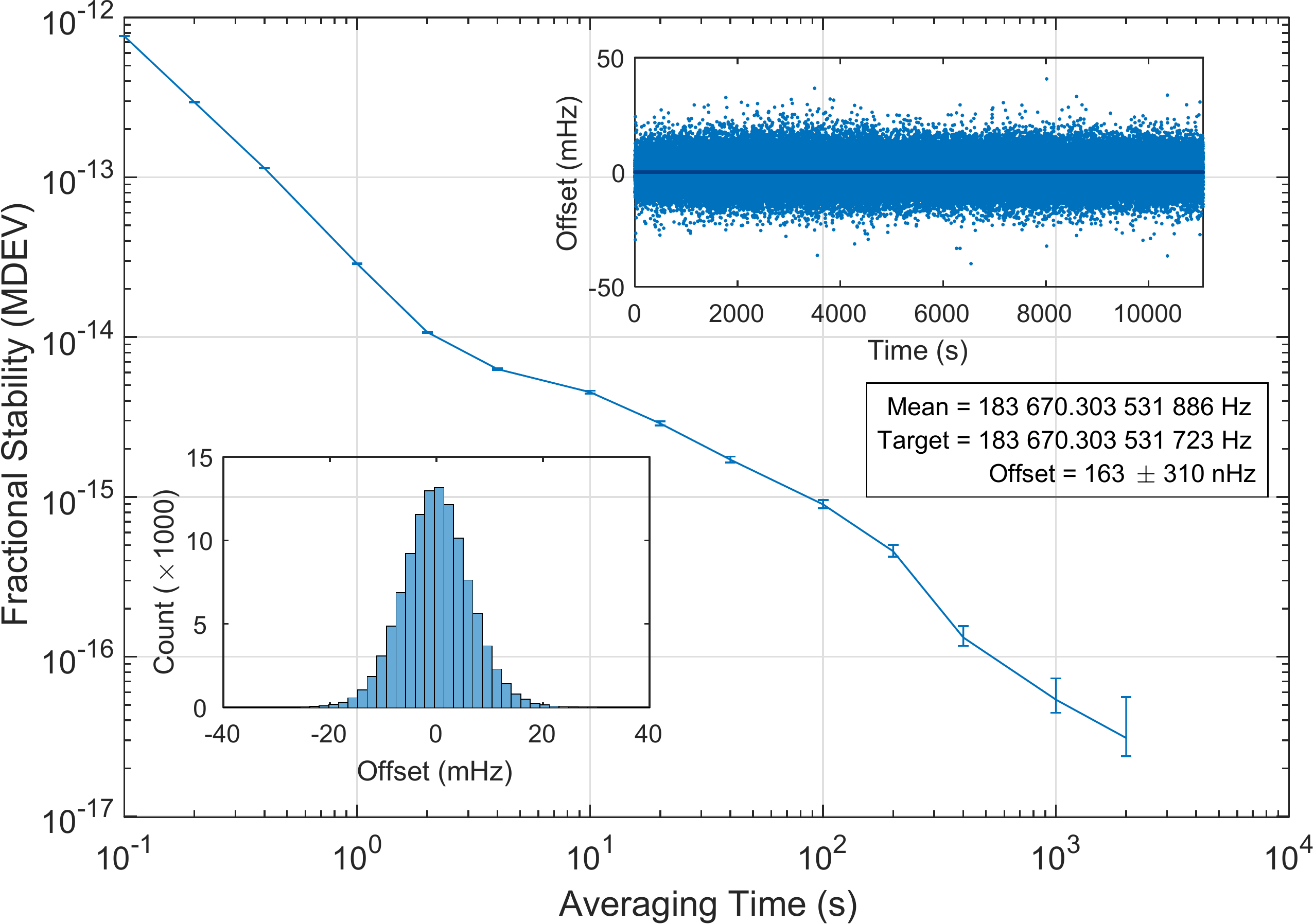}
\caption{Frequency accuracy and stability.  a) Modified Allan Deviation (MDEV) showing the out-of-loop frequency stability of the EOM comb $f_{\rm{rep}}$ measured against a separate self-referenced fiber comb. Two-sided 68\% confidence intervals at each $\tau$ are calculated using $\chi^2$ statistics~\cite{greenhall_estimating_1995}. The top inset shows 11,000~s of frequency-counter data from which the MDEV curve was calculated (expected frequency subtracted) whereas the bottom inset shows a histogram of the data.  The minimum MDEV uncertainty of 3.1$\times 10^{-17}$ at $\tau$~=~2,000~s implies there is no statistical offset in the frequency synthesis of the comb.}
\label{fig:accuracy}
\end{figure}

Fig.~\ref{fig:accuracy} shows the measured stability of the difference in comb repetition rates after an acquisition time of 11,000 s. A zero-dead-time counter ($\Pi$-type) is used to record the RF beat between the two 10~GHz signals while the Modified Allan Deviation (MDEV) is used for the stability calculations. At short averaging times, $\tau < 2$~s, the measured stability is limited by the counter and the $\tau^{-3/2}$ slope is indicative of white phase noise, as expected for two mutually phase locked lasers. For $10 < \tau < 100$~s, differential noise in the system frustrates the phase-coherent averaging, though this noise could be reduced by path-length cancellation of the optical paths, allowing the same degree of stability to be reached at shorter averaging times. At 2,000 s of averaging time, the MDEV yields a minimum uncertainty of 3.1$\times 10^{-17}$ (310~nHz) for a carrier frequency of 10~GHz.  Therefore the measured 163-nHz frequency offset in the mean of the acquisition is statistically consistent with zero synthesis error at the demonstrated level of stability. This level of accuracy represents an improvement of more than three orders of magnitude over previously demonstrated EOM-comb systems~\cite{beha_electronic_2017} and is likely only limited here by averaging time and out-of-loop path differences between the two combs. 

Due to their deterministic nature and fiber-based construction, EOM combs are well suited for sustained maintenance-free operation. To demonstrate this in our system, we run the system continuously for more than 16~hrs and achieve absolute frequency synthesis by fully locking the comb to a hydrogen maser (see Supplementary Fig.~\ref{fig:overnight}).  No glitches in any of the locks are observed within this period, indicating that neither temperature drifts nor beam-pointing fluctuations due to coupling watt-level powers to the SiN chip pose a significant obstacle for continuous operation. This combination of operational reliability and absolute accuracy may make the EOM comb an attractive tool for future long term measurements of ultra-stable optical frequencies and optical clock networks~\cite{riehle_optical_2017}.

\section*{Conclusions}
By effectively eliminating the problems associated with multiplicative noise, while also delivering flat broadband spectra, the promise of few-cycle pulses, and accurate frequency synthesis, our EOM-comb system provides a versatile new ultrafast source. Furthermore, by overcoming several experimental challenges related to broadening and stabilizing noisy picosecond-duration pulses, our techniques are widely applicable to existing technologies with demanding requirements, such as chip-based microresonators or semiconductor lasers.

We note that while resonant elements in the EOM comb currently limit broad tunability of the repetition rate, the comb modes can still be continuously scanned by more than their 10-GHz spacing via tuning of the pump-laser wavelength. This would allow a comb tooth to be set to any arbitrary frequency across the entire multi-octave range of the comb. While this is sufficient for most potential applications, even greater tunability of the repetition rate should be achievable through the use of a lower-noise tunable microwave source that would not require additional noise reduction, possibly in combination with cryogenic cooling.

Finally, the system offers several additional practical benefits.  For instance, these combs may also support more photonic integration through CMOS-compatible modulators \cite{reed_silicon_2010}, alignment-free construction, the use of commercially sourceable components, and straightforward user customization. In the future, we anticipate compact, robust EOM combs to become commonplace across science and technology, enabling new applications for ultrastable femtosecond pulses.

\section*{Acknowledgments}
The authors thank Henry Timmers for assistance making the FROG measurements, Archita Hati and Craig Nelson for discussions on microwave stabilization, Lin Chang for helping lay out the waveguide lithography masks, KV Reddy and Sreenivas Patil for discussion of erbium amplifiers, and Fred Baynes for constructing the cavity-stabilized laser.

This research is supported by the Air Force Office of Scientific Research
(AFOSR) under award number FA9550-16-1-0016, the Defense Advanced Research
Projects Agency (DARPA) DODOS program, the National Aeronautics and Space
Administration (NASA), the National Institute of Standards and Technology
(NIST), and the National Research Council (NRC).

This work is a contribution of the U.S. government and is not subject to
copyright in the U.S.A.

\section*{Author Contributions}
The experiment was planned by D.R.C, D.D.H, S.A.D, and S.B.P. The combs were operated by D.R.C and D.D.H. The optical filter cavity was designed and constructed by W.Z. The 30~GHz measurements were assisted by A.J.M.  The data was analyzed by D.R.C, D.D.H, F.Q., S.A.D., and S.B.P.  The manuscript was prepared by D.R.C. with input from all co-authors.

\bibliographystyle{naturemag}
\bibliography{Zotero}

\begin{thebibliography}{1}
\expandafter\ifx\csname url\endcsname\relax
  \def\url#1{\texttt{#1}}\fi
\expandafter\ifx\csname urlprefix\endcsname\relax\def\urlprefix{URL }\fi
\providecommand{\bibinfo}[2]{#2}
\providecommand{\eprint}[2][]{\url{#2}}

\bibitem{pfeiffer_photonic_2016}
\bibinfo{author}{Pfeiffer, M. H.~P.} \emph{et~al.}
\newblock
  \bibinfo{title}{\href{https://www.osapublishing.org/abstract.cfm?URI=optica-3-1-20}{Photonic
  {Damascene} process for integrated high-{Q} microresonator based nonlinear
  photonics}}.
\newblock \emph{\bibinfo{journal}{Optica}} \textbf{\bibinfo{volume}{3}},
  \bibinfo{pages}{20} (\bibinfo{year}{2016}).

\bibitem{carlson_photonic-chip_2017}
\bibinfo{author}{Carlson, D.~R.} \emph{et~al.}
\newblock
  \bibinfo{title}{\href{http://link.aps.org/doi/10.1103/PhysRevApplied.8.014027}{Photonic-{Chip}
  {Supercontinuum} with {Tailored} {Spectra} for {Counting} {Optical}
  {Frequencies}}}.
\newblock \emph{\bibinfo{journal}{Physical Review Applied}}
  \textbf{\bibinfo{volume}{8}}, \bibinfo{pages}{014027} (\bibinfo{year}{2017}).

\end{thebibliography}


\begin{thebibliography}{10}
\expandafter\ifx\csname url\endcsname\relax
  \def\url#1{\texttt{#1}}\fi
\expandafter\ifx\csname urlprefix\endcsname\relax\def\urlprefix{URL }\fi
\providecommand{\bibinfo}[2]{#2}
\providecommand{\eprint}[2][]{\url{#2}}

\bibitem{cundiff_colloquium:_2003}
\bibinfo{author}{Cundiff, S.~T.} \& \bibinfo{author}{Ye, J.}
\newblock
  \bibinfo{title}{\href{http://link.aps.org/doi/10.1103/RevModPhys.75.325}{Colloquium:
  {Femtosecond} optical frequency combs}}.
\newblock \emph{\bibinfo{journal}{Reviews of Modern Physics}}
  \textbf{\bibinfo{volume}{75}}, \bibinfo{pages}{325--342}
  (\bibinfo{year}{2003}).

\bibitem{kobayashi_highrepetitionrate_1972}
\bibinfo{author}{Kobayashi, T.}, \bibinfo{author}{Sueta, T.},
  \bibinfo{author}{Cho, Y.} \& \bibinfo{author}{Matsuo, Y.}
\newblock
  \bibinfo{title}{\href{http://aip.scitation.org/doi/10.1063/1.1654403}{High‐repetition‐rate
  optical pulse generator using a {Fabry}‐{Perot} electro‐optic
  modulator}}.
\newblock \emph{\bibinfo{journal}{Applied Physics Letters}}
  \textbf{\bibinfo{volume}{21}}, \bibinfo{pages}{341--343}
  (\bibinfo{year}{1972}).

\bibitem{kourogi_wide-span_1993}
\bibinfo{author}{Kourogi, M.}, \bibinfo{author}{Nakagawa, K.} \&
  \bibinfo{author}{Ohtsu, M.}
\newblock
  \bibinfo{title}{\href{http://ieeexplore.ieee.org/abstract/document/250392/}{Wide-span
  optical frequency comb generator for accurate optical frequency difference
  measurement}}.
\newblock \emph{\bibinfo{journal}{IEEE Journal of Quantum Electronics}}
  \textbf{\bibinfo{volume}{29}}, \bibinfo{pages}{2693--2701}
  (\bibinfo{year}{1993}).

\bibitem{torres-company_optical_2014}
\bibinfo{author}{Torres-Company, V.} \& \bibinfo{author}{Weiner, A.~M.}
\newblock
  \bibinfo{title}{\href{http://doi.wiley.com/10.1002/lpor.201300126}{Optical
  frequency comb technology for ultra-broadband radio-frequency photonics:
  {Optical} frequency comb technology for {RF} photonics}}.
\newblock \emph{\bibinfo{journal}{Laser \& Photonics Reviews}}
  \textbf{\bibinfo{volume}{8}}, \bibinfo{pages}{368--393}
  (\bibinfo{year}{2014}).

\bibitem{li_electro-optical_2014}
\bibinfo{author}{Li, J.}, \bibinfo{author}{Yi, X.}, \bibinfo{author}{Lee, H.},
  \bibinfo{author}{Diddams, S.~A.} \& \bibinfo{author}{Vahala, K.~J.}
\newblock
  \bibinfo{title}{\href{http://science.sciencemag.org/content/345/6194/309.abstract}{Electro-optical
  frequency division and stable microwave synthesis}}.
\newblock \emph{\bibinfo{journal}{Science}} \textbf{\bibinfo{volume}{345}},
  \bibinfo{pages}{309} (\bibinfo{year}{2014}).

\bibitem{millot_frequency-agile_2015}
\bibinfo{author}{Millot, G.} \emph{et~al.}
\newblock
  \bibinfo{title}{\href{http://www.nature.com/doifinder/10.1038/nphoton.2015.250}{Frequency-agile
  dual-comb spectroscopy}}.
\newblock \emph{\bibinfo{journal}{Nature Photonics}}
  \textbf{\bibinfo{volume}{10}}, \bibinfo{pages}{27--30}
  (\bibinfo{year}{2015}).

\bibitem{beha_electronic_2017}
\bibinfo{author}{Beha, K.} \emph{et~al.}
\newblock
  \bibinfo{title}{\href{https://www.osapublishing.org/abstract.cfm?URI=optica-4-4-406}{Electronic
  synthesis of light}}.
\newblock \emph{\bibinfo{journal}{Optica}} \textbf{\bibinfo{volume}{4}},
  \bibinfo{pages}{406} (\bibinfo{year}{2017}).

\bibitem{herr_temporal_2013}
\bibinfo{author}{Herr, T.} \emph{et~al.}
\newblock
  \bibinfo{title}{\href{http://www.nature.com/doifinder/10.1038/nphoton.2013.343}{Temporal
  solitons in optical microresonators}}.
\newblock \emph{\bibinfo{journal}{Nature Photonics}}
  \textbf{\bibinfo{volume}{8}}, \bibinfo{pages}{145--152}
  (\bibinfo{year}{2013}).

\bibitem{tilma_recent_2015}
\bibinfo{author}{Tilma, B.~W.} \emph{et~al.}
\newblock
  \bibinfo{title}{\href{http://www.nature.com/doifinder/10.1038/lsa.2015.83}{Recent
  advances in ultrafast semiconductor disk lasers}}.
\newblock \emph{\bibinfo{journal}{Light: Science \& Applications}}
  \textbf{\bibinfo{volume}{4}}, \bibinfo{pages}{e310} (\bibinfo{year}{2015}).

\bibitem{marin-palomo_microresonator-based_2017}
\bibinfo{author}{Marin-Palomo, P.} \emph{et~al.}
\newblock
  \bibinfo{title}{\href{http://www.nature.com/doifinder/10.1038/nature22387}{Microresonator-based
  solitons for massively parallel coherent optical communications}}.
\newblock \emph{\bibinfo{journal}{Nature}} \textbf{\bibinfo{volume}{546}},
  \bibinfo{pages}{274--279} (\bibinfo{year}{2017}).

\bibitem{li_laser_2008}
\bibinfo{author}{Li, C.-H.} \emph{et~al.}
\newblock
  \bibinfo{title}{\href{http://www.nature.com/doifinder/10.1038/nature06854}{A
  laser frequency comb that enables radial velocity measurements with a
  precision of 1 cm s⁻¹}}.
\newblock \emph{\bibinfo{journal}{Nature}} \textbf{\bibinfo{volume}{452}},
  \bibinfo{pages}{610--612} (\bibinfo{year}{2008}).

\bibitem{steinmetz_laser_2008}
\bibinfo{author}{Steinmetz, T.} \emph{et~al.}
\newblock
  \bibinfo{title}{\href{http://www.sciencemag.org/content/321/5894/1335}{Laser
  {Frequency} {Combs} for {Astronomical} {Observations}}}.
\newblock \emph{\bibinfo{journal}{Science}} \textbf{\bibinfo{volume}{321}},
  \bibinfo{pages}{1335--1337} (\bibinfo{year}{2008}).

\bibitem{camp_jr_chemically_2015}
\bibinfo{author}{Camp~Jr, C.~H.} \& \bibinfo{author}{Cicerone, M.~T.}
\newblock
  \bibinfo{title}{\href{http://www.nature.com/doifinder/10.1038/nphoton.2015.60}{Chemically
  sensitive bioimaging with coherent {Raman} scattering}}.
\newblock \emph{\bibinfo{journal}{Nature Photonics}}
  \textbf{\bibinfo{volume}{9}}, \bibinfo{pages}{295--305}
  (\bibinfo{year}{2015}).

\bibitem{coddington_dual-comb_2016}
\bibinfo{author}{Coddington, I.}, \bibinfo{author}{Newbury, N.} \&
  \bibinfo{author}{Swann, W.}
\newblock
  \bibinfo{title}{\href{https://www.osapublishing.org/abstract.cfm?URI=optica-3-4-414}{Dual-comb
  spectroscopy}}.
\newblock \emph{\bibinfo{journal}{Optica}} \textbf{\bibinfo{volume}{3}},
  \bibinfo{pages}{414} (\bibinfo{year}{2016}).

\bibitem{heinecke_optical_2009}
\bibinfo{author}{Heinecke, D.~C.} \emph{et~al.}
\newblock
  \bibinfo{title}{\href{https://link.aps.org/doi/10.1103/PhysRevA.80.053806}{Optical
  frequency stabilization of a 10 {GHz} {Ti}:sapphire frequency comb by
  saturated absorption spectroscopy in ⁸⁷rubidium}}.
\newblock \emph{\bibinfo{journal}{Physical Review A}}
  \textbf{\bibinfo{volume}{80}} (\bibinfo{year}{2009}).

\bibitem{duran_electro-optic_2016}
\bibinfo{author}{Durán, V.}, \bibinfo{author}{Andrekson, P.~A.} \&
  \bibinfo{author}{Torres-Company, V.}
\newblock
  \bibinfo{title}{\href{https://www.osapublishing.org/abstract.cfm?URI=ol-41-18-4190}{Electro-optic
  dual-comb interferometry over 40 nm bandwidth}}.
\newblock \emph{\bibinfo{journal}{Optics Letters}}
  \textbf{\bibinfo{volume}{41}}, \bibinfo{pages}{4190} (\bibinfo{year}{2016}).

\bibitem{kobayashi_optical_1988}
\bibinfo{author}{Kobayashi, T.} \emph{et~al.}
\newblock
  \bibinfo{title}{\href{http://ieeexplore.ieee.org/xpls/abs_all.jsp?arnumber=135}{Optical
  pulse compression using high-frequency electrooptic phase modulation}}.
\newblock \emph{\bibinfo{journal}{IEEE Journal of Quantum Electronics}}
  \textbf{\bibinfo{volume}{24}}, \bibinfo{pages}{382--387}
  (\bibinfo{year}{1988}).

\bibitem{metcalf_broadly_2015}
\bibinfo{author}{Metcalf, A.~J.}, \bibinfo{author}{Quinlan, F.},
  \bibinfo{author}{Fortier, T.~M.}, \bibinfo{author}{Diddams, S.~A.} \&
  \bibinfo{author}{Weiner, A.~M.}
\newblock
  \bibinfo{title}{\href{http://digital-library.theiet.org/content/journals/10.1049/el.2015.1367}{Broadly
  tunable, low timing jitter, high repetition rate optoelectronic comb
  generator}}.
\newblock \emph{\bibinfo{journal}{Electronics Letters}}
  \textbf{\bibinfo{volume}{51}}, \bibinfo{pages}{1596--1598}
  (\bibinfo{year}{2015}).

\bibitem{walls_rf_1975}
\bibinfo{author}{Walls, F.~L.} \& \bibinfo{author}{DeMarchi, A.}
\newblock
  \bibinfo{title}{\href{http://ieeexplore.ieee.org/abstract/document/4314411/}{{RF}
  spectrum of a signal after frequency multiplication; measurement and
  comparison with a simple calculation}}.
\newblock \emph{\bibinfo{journal}{IEEE Transactions on Instrumentation and
  Measurement}} \textbf{\bibinfo{volume}{24}}, \bibinfo{pages}{210--217}
  (\bibinfo{year}{1975}).

\bibitem{kim_cavity-induced_2017}
\bibinfo{author}{Kim, J.}, \bibinfo{author}{Richardson, D.~J.} \&
  \bibinfo{author}{Slavík, R.}
\newblock
  \bibinfo{title}{\href{https://www.osapublishing.org/abstract.cfm?URI=ol-42-8-1536}{Cavity-induced
  phase noise suppression in a {Fabry}–{Perot} modulator-based optical
  frequency comb}}.
\newblock \emph{\bibinfo{journal}{Optics Letters}}
  \textbf{\bibinfo{volume}{42}}, \bibinfo{pages}{1536} (\bibinfo{year}{2017}).

\bibitem{dudley_supercontinuum_2006}
\bibinfo{author}{Dudley, J.~M.}, \bibinfo{author}{Genty, G.} \&
  \bibinfo{author}{Coen, S.}
\newblock
  \bibinfo{title}{\href{http://link.aps.org/doi/10.1103/RevModPhys.78.1135}{Supercontinuum
  generation in photonic crystal fiber}}.
\newblock \emph{\bibinfo{journal}{Reviews of Modern Physics}}
  \textbf{\bibinfo{volume}{78}}, \bibinfo{pages}{1135--1184}
  (\bibinfo{year}{2006}).

\bibitem{tamura_fundamentals_2000}
\bibinfo{author}{Tamura, K.~R.}, \bibinfo{author}{Kuhota, H.} \&
  \bibinfo{author}{Nakazawa, M.}
\newblock
  \bibinfo{title}{\href{http://ieeexplore.ieee.org/abstract/document/848347/}{Fundamentals
  of stable continuum generation at high repetition rates}}.
\newblock \emph{\bibinfo{journal}{IEEE Journal of Quantum Electronics}}
  \textbf{\bibinfo{volume}{36}}, \bibinfo{pages}{773--779}
  (\bibinfo{year}{2000}).

\bibitem{huang_nonlinearly_2008}
\bibinfo{author}{Huang, C.-B.}, \bibinfo{author}{Park, S.-G.},
  \bibinfo{author}{Leaird, D.~E.} \& \bibinfo{author}{Weiner, A.~M.}
\newblock
  \bibinfo{title}{\href{http://www.opticsexpress.org/abstract.cfm?URI=oe-16-4-2520}{Nonlinearly
  broadened phase-modulated continuous-wave laser frequency combs characterized
  using {DPSK} decoding}}.
\newblock \emph{\bibinfo{journal}{Optics Express}}
  \textbf{\bibinfo{volume}{16}}, \bibinfo{pages}{2520--2527}
  (\bibinfo{year}{2008}).

\bibitem{cole_octave-spanning_2016}
\bibinfo{author}{Cole, D.~C.}, \bibinfo{author}{Beha, K.~M.},
  \bibinfo{author}{Diddams, S.~A.} \& \bibinfo{author}{Papp, S.~B.}
\newblock
  \bibinfo{title}{\href{http://stacks.iop.org/1742-6596/723/i=1/a=012035?key=crossref.09b6ad4818b90f7776c58180de3fafc0}{Octave-spanning
  supercontinuum generation via microwave frequency multiplication}}.
\newblock \emph{\bibinfo{journal}{Journal of Physics: Conference Series}}
  \textbf{\bibinfo{volume}{723}}, \bibinfo{pages}{012035}
  (\bibinfo{year}{2016}).

\bibitem{wu_supercontinuum-based_2013}
\bibinfo{author}{Wu, R.}, \bibinfo{author}{Torres-Company, V.},
  \bibinfo{author}{Leaird, D.~E.} \& \bibinfo{author}{Weiner, A.~M.}
\newblock
  \bibinfo{title}{\href{http://www.opticsexpress.org/abstract.cfm?URI=oe-21-5-6045}{Supercontinuum-based
  10-{GHz} flat-topped optical frequency comb generation}}.
\newblock \emph{\bibinfo{journal}{Optics Express}}
  \textbf{\bibinfo{volume}{21}}, \bibinfo{pages}{6045--6052}
  (\bibinfo{year}{2013}).

\bibitem{bartels_10-ghz_2009}
\bibinfo{author}{Bartels, A.}, \bibinfo{author}{Heinecke, D.} \&
  \bibinfo{author}{Diddams, S.~A.}
\newblock
  \bibinfo{title}{\href{http://www.sciencemag.org/cgi/doi/10.1126/science.1179112}{10-{GHz}
  {Self}-{Referenced} {Optical} {Frequency} {Comb}}}.
\newblock \emph{\bibinfo{journal}{Science}} \textbf{\bibinfo{volume}{326}},
  \bibinfo{pages}{681--681} (\bibinfo{year}{2009}).

\bibitem{kashiwagi_direct_2016}
\bibinfo{author}{Kashiwagi, K.} \emph{et~al.}
\newblock
  \bibinfo{title}{\href{https://www.osapublishing.org/abstract.cfm?URI=oe-24-8-8120}{Direct
  generation of 12.5-{GHz}-spaced optical frequency comb with ultrabroad
  coverage in near-infrared region by cascaded fiber configuration}}.
\newblock \emph{\bibinfo{journal}{Optics Express}}
  \textbf{\bibinfo{volume}{24}}, \bibinfo{pages}{8120} (\bibinfo{year}{2016}).

\bibitem{johnson_octave-spanning_2015}
\bibinfo{author}{Johnson, A.~R.} \emph{et~al.}
\newblock
  \bibinfo{title}{\href{https://www.osapublishing.org/abstract.cfm?URI=ol-40-21-5117}{Octave-spanning
  coherent supercontinuum generation in a silicon nitride waveguide}}.
\newblock \emph{\bibinfo{journal}{Optics Letters}}
  \textbf{\bibinfo{volume}{40}}, \bibinfo{pages}{5117} (\bibinfo{year}{2015}).

\bibitem{zhao_visible--near-infrared_2015}
\bibinfo{author}{Zhao, H.} \emph{et~al.}
\newblock
  \bibinfo{title}{\href{https://www.osapublishing.org/abstract.cfm?URI=ol-40-10-2177}{Visible-to-near-infrared
  octave spanning supercontinuum generation in a silicon nitride waveguide}}.
\newblock \emph{\bibinfo{journal}{Optics Letters}}
  \textbf{\bibinfo{volume}{40}}, \bibinfo{pages}{2177} (\bibinfo{year}{2015}).

\bibitem{klenner_gigahertz_2016}
\bibinfo{author}{Klenner, A.} \emph{et~al.}
\newblock
  \bibinfo{title}{\href{https://www.osapublishing.org/abstract.cfm?URI=oe-24-10-11043}{Gigahertz
  frequency comb offset stabilization based on supercontinuum generation in
  silicon nitride waveguides}}.
\newblock \emph{\bibinfo{journal}{Optics Express}}
  \textbf{\bibinfo{volume}{24}}, \bibinfo{pages}{11043} (\bibinfo{year}{2016}).

\bibitem{porcel_two-octave_2017}
\bibinfo{author}{Porcel, M. A.~G.} \emph{et~al.}
\newblock
  \bibinfo{title}{\href{https://www.osapublishing.org/abstract.cfm?URI=oe-25-2-1542}{Two-octave
  spanning supercontinuum generation in stoichiometric silicon nitride
  waveguides pumped at telecom wavelengths}}.
\newblock \emph{\bibinfo{journal}{Optics Express}}
  \textbf{\bibinfo{volume}{25}}, \bibinfo{pages}{1542} (\bibinfo{year}{2017}).

\bibitem{carlson_self-referenced_2017}
\bibinfo{author}{Carlson, D.~R.} \emph{et~al.}
\newblock
  \bibinfo{title}{\href{https://www.osapublishing.org/abstract.cfm?URI=ol-42-12-2314}{Self-referenced
  frequency combs using high-efficiency silicon-nitride waveguides}}.
\newblock \emph{\bibinfo{journal}{Optics Letters}}
  \textbf{\bibinfo{volume}{42}}, \bibinfo{pages}{2314} (\bibinfo{year}{2017}).

\bibitem{hickstein_ultrabroadband_2017}
\bibinfo{author}{Hickstein, D.~D.} \emph{et~al.}
\newblock
  \bibinfo{title}{\href{http://link.aps.org/doi/10.1103/PhysRevApplied.8.014025}{Ultrabroadband
  {Supercontinuum} {Generation} and {Frequency}-{Comb} {Stabilization} {Using}
  {On}-{Chip} {Waveguides} with {Both} {Cubic} and {Quadratic}
  {Nonlinearities}}}.
\newblock \emph{\bibinfo{journal}{Physical Review Applied}}
  \textbf{\bibinfo{volume}{8}}, \bibinfo{pages}{014025} (\bibinfo{year}{2017}).

\bibitem{braje_astronomical_2008}
\bibinfo{author}{Braje, D.~A.}, \bibinfo{author}{Kirchner, M.~S.},
  \bibinfo{author}{Osterman, S.}, \bibinfo{author}{Fortier, T.} \&
  \bibinfo{author}{Diddams, S.~A.}
\newblock
  \bibinfo{title}{\href{http://www.springerlink.com/index/10.1140/epjd/e2008-00099-9}{Astronomical
  spectrograph calibration with broad-spectrum frequency combs}}.
\newblock \emph{\bibinfo{journal}{The European Physical Journal D}}
  \textbf{\bibinfo{volume}{48}}, \bibinfo{pages}{57--66}
  (\bibinfo{year}{2008}).

\bibitem{steinmetz_fabryperot_2009}
\bibinfo{author}{Steinmetz, T.} \emph{et~al.}
\newblock
  \bibinfo{title}{\href{http://link.springer.com/10.1007/s00340-009-3374-6}{Fabry–{Pérot}
  filter cavities for wide-spaced frequency combs with large spectral
  bandwidth}}.
\newblock \emph{\bibinfo{journal}{Applied Physics B}}
  \textbf{\bibinfo{volume}{96}}, \bibinfo{pages}{251--256}
  (\bibinfo{year}{2009}).

\bibitem{ycas_demonstration_2012}
\bibinfo{author}{Ycas, G.~G.} \emph{et~al.}
\newblock
  \bibinfo{title}{\href{http://www.opticsexpress.org/abstract.cfm?URI=oe-20-6-6631}{Demonstration
  of on-sky calibration of astronomical spectra using a 25 {GHz} near-{IR}
  laser frequency comb}}.
\newblock \emph{\bibinfo{journal}{Optics Express}}
  \textbf{\bibinfo{volume}{20}}, \bibinfo{pages}{6631--6643}
  (\bibinfo{year}{2012}).

\bibitem{mccracken_decade_2017}
\bibinfo{author}{McCracken, R.~A.}, \bibinfo{author}{Charsley, J.~M.} \&
  \bibinfo{author}{Reid, D.~T.}
\newblock
  \bibinfo{title}{\href{https://www.osapublishing.org/abstract.cfm?URI=oe-25-13-15058}{Decade
  of astrocombs: recent advances in frequency combs for astronomy}}.
\newblock \emph{\bibinfo{journal}{Optics Express}}
  \textbf{\bibinfo{volume}{25}}, \bibinfo{pages}{15058} (\bibinfo{year}{2017}).

\bibitem{jones_carrier-envelope_2000}
\bibinfo{author}{Jones, D.~J.} \emph{et~al.}
\newblock
  \bibinfo{title}{\href{http://www.sciencemag.org/content/288/5466/635}{Carrier-{Envelope}
  {Phase} {Control} of {Femtosecond} {Mode}-{Locked} {Lasers} and {Direct}
  {Optical} {Frequency} {Synthesis}}}.
\newblock \emph{\bibinfo{journal}{Science}} \textbf{\bibinfo{volume}{288}},
  \bibinfo{pages}{635--639} (\bibinfo{year}{2000}).

\bibitem{dick_measurement_1989}
\bibinfo{author}{Dick, G.~J.} \& \bibinfo{author}{Saunders, J.}
\newblock
  \bibinfo{title}{\href{http://ieeexplore.ieee.org/document/68843/}{Measurement
  and analysis of a microwave oscillator stabilized by a sapphire dielectric
  ring resonator for ultra-low noise}}.
\newblock In \emph{\bibinfo{booktitle}{Proceedings of the 43rd {Annual}
  {Symposium} on {Frequency} {Control}}}, \bibinfo{pages}{107--114}
  (\bibinfo{year}{1989}).

\bibitem{gupta_high_2004}
\bibinfo{author}{Gupta, A.~S.} \emph{et~al.}
\newblock
  \bibinfo{title}{\href{http://ieeexplore.ieee.org/abstract/document/1350949/}{High
  spectral purity microwave oscillator: {Design} using conventional
  air-dielectric cavity}}.
\newblock \emph{\bibinfo{journal}{IEEE Transactions on Ultrasonics,
  Ferroelectrics, and Frequency Control}} \textbf{\bibinfo{volume}{51}},
  \bibinfo{pages}{1225--1231} (\bibinfo{year}{2004}).

\bibitem{di_domenico_simple_2010}
\bibinfo{author}{Di~Domenico, G.}, \bibinfo{author}{Schilt, S.} \&
  \bibinfo{author}{Thomann, P.}
\newblock
  \bibinfo{title}{\href{https://www.osapublishing.org/abstract.cfm?uri=ao-49-25-4801}{Simple
  approach to the relation between laser frequency noise and laser line
  shape}}.
\newblock \emph{\bibinfo{journal}{Applied Optics}}
  \textbf{\bibinfo{volume}{49}}, \bibinfo{pages}{4801--4807}
  (\bibinfo{year}{2010}).

\bibitem{jiang_spectral_2007}
\bibinfo{author}{Jiang, Z.} \emph{et~al.}
\newblock
  \bibinfo{title}{\href{http://ieeexplore.ieee.org/document/4376270/}{Spectral
  {Line}-by-{Line} {Pulse} {Shaping} on an {Optical} {Frequency} {Comb}
  {Generator}}}.
\newblock \emph{\bibinfo{journal}{IEEE Journal of Quantum Electronics}}
  \textbf{\bibinfo{volume}{43}}, \bibinfo{pages}{1163--1174}
  (\bibinfo{year}{2007}).

\bibitem{zewail_femtochemistry:_2000}
\bibinfo{author}{Zewail, A.~H.}
\newblock
  \bibinfo{title}{\href{http://pubs.acs.org/doi/abs/10.1021/jp001460h}{Femtochemistry:
  {Atomic}-{Scale} {Dynamics} of the {Chemical} {Bond}}}.
\newblock \emph{\bibinfo{journal}{The Journal of Physical Chemistry A}}
  \textbf{\bibinfo{volume}{104}}, \bibinfo{pages}{5660--5694}
  (\bibinfo{year}{2000}).

\bibitem{corkum_attosecond_2007}
\bibinfo{author}{Corkum, P.~B.} \& \bibinfo{author}{Krausz, F.}
\newblock
  \bibinfo{title}{\href{http://www.nature.com/nphys/journal/v3/n6/full/nphys620.html}{Attosecond
  science}}.
\newblock \emph{\bibinfo{journal}{Nature Physics}}
  \textbf{\bibinfo{volume}{3}}, \bibinfo{pages}{381} (\bibinfo{year}{2007}).

\bibitem{gattass_femtosecond_2008}
\bibinfo{author}{Gattass, R.~R.} \& \bibinfo{author}{Mazur, E.}
\newblock
  \bibinfo{title}{\href{https://www.nature.com/articles/nphoton.2008.47}{Femtosecond
  laser micromachining in transparent materials}}.
\newblock \emph{\bibinfo{journal}{Nature Photonics}}
  \textbf{\bibinfo{volume}{2}}, \bibinfo{pages}{219--225}
  (\bibinfo{year}{2008}).

\bibitem{brabec_intense_2000}
\bibinfo{author}{Brabec, T.} \& \bibinfo{author}{Krausz, F.}
\newblock
  \bibinfo{title}{\href{https://journals.aps.org/rmp/abstract/10.1103/RevModPhys.72.545}{Intense
  few-cycle laser fields: {Frontiers} of nonlinear optics}}.
\newblock \emph{\bibinfo{journal}{Reviews of Modern Physics}}
  \textbf{\bibinfo{volume}{72}}, \bibinfo{pages}{545} (\bibinfo{year}{2000}).

\bibitem{ideguchi_coherent_2013}
\bibinfo{author}{Ideguchi, T.} \emph{et~al.}
\newblock
  \bibinfo{title}{\href{http://www.nature.com/doifinder/10.1038/nature12607}{Coherent
  {Raman} spectro-imaging with laser frequency combs}}.
\newblock \emph{\bibinfo{journal}{Nature}} \textbf{\bibinfo{volume}{502}},
  \bibinfo{pages}{355--358} (\bibinfo{year}{2013}).

\bibitem{agrawal_nonlinear_2013}
\bibinfo{author}{Agrawal, G.~P.}
\newblock \emph{\bibinfo{title}{Nonlinear {Fiber} {Optics}}}
  (\bibinfo{publisher}{Academic Press}, \bibinfo{year}{2013}).

\bibitem{mollenauer_extreme_1983}
\bibinfo{author}{Mollenauer, L.~F.}, \bibinfo{author}{Tomlinson, W.~J.},
  \bibinfo{author}{Stolen, R.~H.} \& \bibinfo{author}{Gordon, J.~P.}
\newblock
  \bibinfo{title}{\href{https://www.osapublishing.org/abstract.cfm?uri=ol-8-5-289}{Extreme
  picosecond pulse narrowing by means of soliton effect in single-mode optical
  fibers}}.
\newblock \emph{\bibinfo{journal}{Optics Letters}}
  \textbf{\bibinfo{volume}{8}}, \bibinfo{pages}{289--291}
  (\bibinfo{year}{1983}).

\bibitem{trebino_measuring_1997}
\bibinfo{author}{Trebino, R.} \emph{et~al.}
\newblock
  \bibinfo{title}{\href{http://aip.scitation.org/doi/10.1063/1.1148286}{Measuring
  ultrashort laser pulses in the time-frequency domain using frequency-resolved
  optical gating}}.
\newblock \emph{\bibinfo{journal}{Review of Scientific Instruments}}
  \textbf{\bibinfo{volume}{68}}, \bibinfo{pages}{3277--3295}
  (\bibinfo{year}{1997}).

\bibitem{sidorenko_ptychographic_2016}
\bibinfo{author}{Sidorenko, P.}, \bibinfo{author}{Lahav, O.},
  \bibinfo{author}{Avnat, Z.} \& \bibinfo{author}{Cohen, O.}
\newblock
  \bibinfo{title}{\href{https://www.osapublishing.org/abstract.cfm?URI=optica-3-12-1320}{Ptychographic
  reconstruction algorithm for frequency-resolved optical gating:
  super-resolution and supreme robustness}}.
\newblock \emph{\bibinfo{journal}{Optica}} \textbf{\bibinfo{volume}{3}},
  \bibinfo{pages}{1320} (\bibinfo{year}{2016}).

\bibitem{greenhall_estimating_1995}
\bibinfo{author}{Greenhall, C.~A.}
\newblock
  \bibinfo{title}{\href{http://ieeexplore.ieee.org/document/483920/}{Estimating
  the modified {Allan} variance}}.
\newblock In \emph{\bibinfo{booktitle}{Proceedings of the 1995 {IEEE}
  {International} {Frequency} {Control} {Symposium}}},
  \bibinfo{pages}{346--353} (\bibinfo{year}{1995}).

\bibitem{riehle_optical_2017}
\bibinfo{author}{Riehle, F.}
\newblock
  \bibinfo{title}{\href{http://www.nature.com/doifinder/10.1038/nphoton.2016.235}{Optical
  clock networks}}.
\newblock \emph{\bibinfo{journal}{Nature Photonics}}
  \textbf{\bibinfo{volume}{11}}, \bibinfo{pages}{25--31}
  (\bibinfo{year}{2017}).

\bibitem{reed_silicon_2010}
\bibinfo{author}{Reed, G.~T.}, \bibinfo{author}{Mashanovich, G.},
  \bibinfo{author}{Gardes, F.~Y.} \& \bibinfo{author}{Thomson, D.~J.}
\newblock
  \bibinfo{title}{\href{http://www.nature.com/doifinder/10.1038/nphoton.2010.179}{Silicon
  optical modulators}}.
\newblock \emph{\bibinfo{journal}{Nature Photonics}}
  \textbf{\bibinfo{volume}{4}}, \bibinfo{pages}{518--526}
  (\bibinfo{year}{2010}).

\end{thebibliography}

\clearpage
\small
\section*{Methods}
\subsection*{Comb generation and microwave stabilization}
The central mode of the EOM comb is formed from a \mbox{1550-nm} CW laser that is stabilized with sub-hertz linewidth to an ultra-low expansion cavity.  For maximal frequency tunability though, the comb can also be operated successfully with an external-cavity diode laser as the pump.  The CW laser is then transmitted through two to four lithium-niobate phase modulators with low half-wave voltage ($V_\pi \approx 3$ V), optically connected in series, to produce a periodically chirped continuous waveform. At the highest drive power of 36~dBm each modulator can make approximately 40 comb lines.  An intensity modulator driven by the same microwave drive then carves the chirped output into a 50\% duty-cycle pulse train.  All components in the comb generator are polarization-maintaining for convenience and reliability.

The aluminum microwave cavity used to pre-stabilize the 10-GHz DRO (Synergy Microwave DRO100) has a cylindrical geometry and is operated in the TE$_{011}$ mode with an unloaded Q-factor of 8300. The cavity was machined in-house, though we note that comparable commercial options are available. Leakage signal from one of the optical phase modulators is coupled to the cavity with an incident power up to 30~dBm.  The resonator is operated in a nearly critically coupled regime in order to achieve the maximum carrier suppression ($>$30~dB) and frequency-discrimination sensitivity from the reflected signal.  An error signal for locking the DRO to the cavity is generated by mixing this reflected signal with the original DRO output.

\subsection*{Nonlinear broadening}
Following the filter cavity, the pulse train is amplified in an erbium-doped fiber amplifier (Pritel SP-LNHP-FA-37-IO-NMA, maximum output power 5~W) before the first stage of spectral broadening in highly nonlinear fiber (HNLF).  The HNLF has normal dispersion at 1550~nm \mbox{($D = -1.4$ ps/nm/km)} and its free-space output facet is angle-cleaved to prevent back reflections.  A pair of high-transmission diffraction gratings (940~grooves/mm) is then used to compensate the HNLF dispersion and deliver sub-100-fs pulses to the SiN waveguide.

The photonic waveguides used for the second-stage broadening were commercially fabricated by Ligentec using the ``Photonic Damascene'' process \citeMethods{pfeiffer_photonic_2016}.  They are made from low-pressure chemical-vapor-deposition (LPCVD) SiN, are 15~mm in length, and have a thickness of 750~nm.  For self-referencing, a waveguide width of 1800~nm is used, though the spectrum is easily adjustable by changing this parameter \citeMethods{carlson_photonic-chip_2017}.  Additionally, the waveguides have a fully oxide-clad (SiO$_2$) geometry that serves to both protect the waveguide from contamination as well as likely improving the ability of the device to withstand sustained thermal loads when incident with up to several watts of optical power.  Input coupling to the waveguide is accomplished using an aspheric lens with a design wavelength of 1550~nm and a numerical aperture (NA) of 0.6 while output-coupling is achieved with a 0.85-NA visible-wavelength microscope objective.  Both facets of the chip include inverse tapers (minimum dimension = 150~nm, length 300~$\mu$m) for improved coupling to achieve a total power throughput greater than 70\%.

Spectral flattening of the waveguide output can be accomplished using a passive in-line fiber attenuator with a fixed 20-dB attenuation coefficient at 1550~nm. An exponential roll-off in the attenuation coefficient in going to shorter wavelengths allows flattening of the spectral amplitude to within a variation of $\pm3$~dB, spanning from 850~nm to 1450~nm.

\subsection*{Offset-frequency stabilization}

A fiber-coupled waveguide PPLN is used for second-harmonic generation of residual 1550-nm pump light (100~mW) exiting the SiN waveguide for $f$--$2f$ detection. The PPLN output is temporally and spatially overlapped with the 775-nm supercontinuum light, coupled to single-mode fiber, and delivered to a 10-GHz-bandwidth photodetector with a built-in trans-impedance amplifier.  The offset frequency is detected near 3.5~GHz and is roughly determined by the filter cavity geometry and mirror coatings. After detection, this beat is electronically mixed down to 960~MHz before filtering and frequency division by 32.  A phase-locked loop locks the resulting 30-MHz signal to a hydrogen-maser-referenced synthesizer by adjusting the voltage set point of the microwave cavity lock.  An auxiliary output of the feedback loop is also used to provide a slow thermal correction to the microwave cavity through resistive heaters affixed to the outside.

\bibliographystyleMethods{naturemag}
\bibliographyMethods{Zotero}

\vspace{5mm}
\small
Note: Certain commercial equipment, instruments, or materials are identified in this
paper in order to specify the experimental procedure adequately. Such
identification is not intended to imply recommendation or endorsement by the
National Institute of Standards and Technology, nor is it intended to imply that
the materials or equipment identified are necessarily the best available for the
purpose.

\clearpage
\normalsize
\renewcommand{\thefigure}{S\arabic{figure}}
\setcounter{figure}{0}   

\onecolumn
\section*{\centering Supplementary Material}
This supplement provides additional experimental details and data to support the claims and conclusions discussed in the main text.

\subsection*{Experimental schematic}

\begin{figure}[h]
\centering
\includegraphics[width=\linewidth]{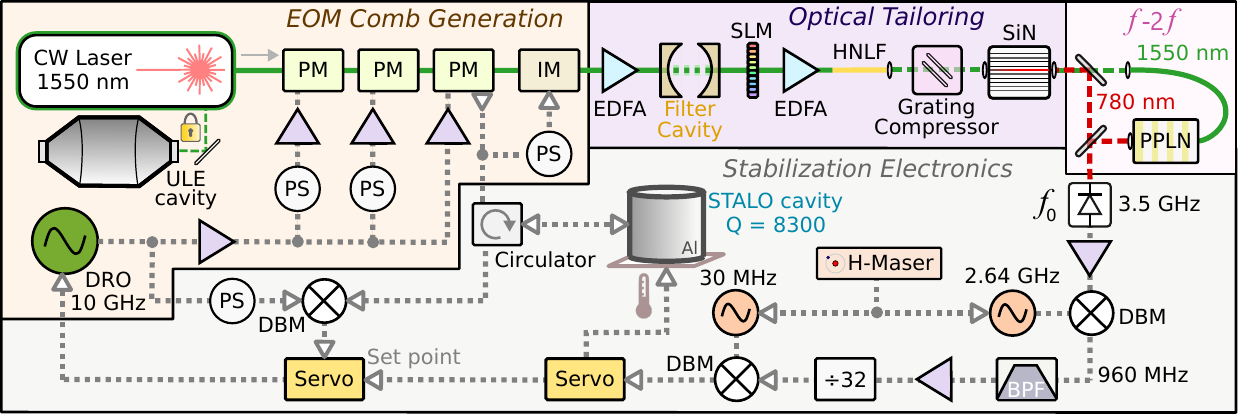}
\caption{Experimental schematic for comb generation and stabilization. See main text for details. Fiber paths, solid lines; free-space paths, dashed lines; electrical, gray dotted lines.  ULE, ultra-low expansion cavity, PM, phase modulation; IM, intensity modulation; DRO, dielectric resonant oscillator; PS, phase shifter; EDFA, erbium-doped fiber amplifier; SLM, spatial light modulator for dispersion control; HNLF, highly nonlinear fiber; SiN, silicon-nitride waveguide; BPF, band-pass filter; DBM, double-balanced mixer; STALO, stabilized-local-oscillator cavity; PPLN, periodically poled lithium niobate.}
\label{fig:schematic}
\end{figure}

A schematic of the 10-GHz EOM comb, including both the optical and microwave stabilization components, is shown in Fig.~\ref{fig:schematic}.  See the main text for details.

\subsection*{Supercontinuum coherence and initial comb bandwidth}
\begin{figure*}
\centering
\includegraphics[width=\linewidth]{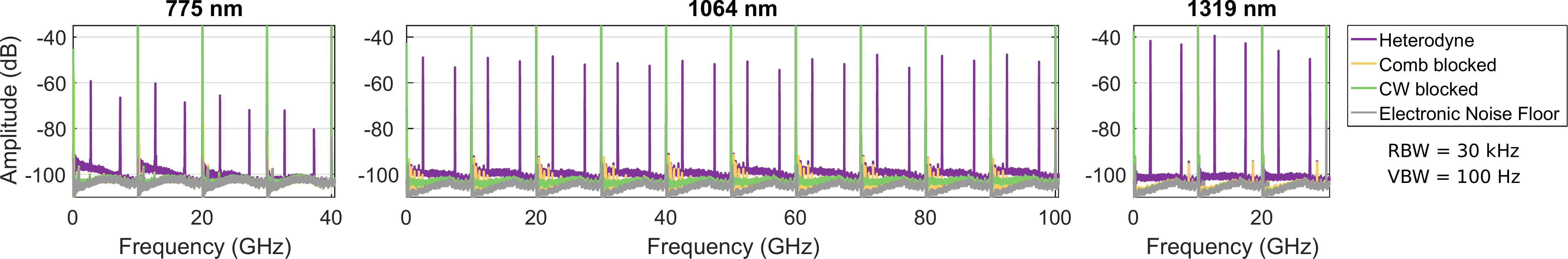}
\caption{High-resolution optical spectrum of the 10-GHz EOM comb obtained from heterodyne beats with CW lasers.  Clean, high-contrast comb modes are obtained across the entire bandwidth of the supercontinuum spectrum.  The fall-off in signal-to-noise ratio in the 775-nm band at higher frequencies is due to reduced efficiency of the doubling crystal used to generate the CW light.}
\label{fig:cwheterodyne}
\end{figure*}

Individual comb lines in the silicon-nitride supercontinuum spectrum exhibit a high degree of extinction across the entire bandwidth, as shown by the CW heterodyne beats in Fig.~\ref{fig:cwheterodyne}.  For each wavelength region shown, the spectra are obtained by stepping the frequency of a CW laser in 10~GHz increments and recording the radio-frequency (RF) heterodyne spectrum at each step.  Bandwidths greater than 10~GHz are recorded by stitching together multiple acquisitions at each wavelength.  Each individual 10~GHz span then contains a beat between the CW laser and the two nearest comb modes.  Individual noise floor contributions from the CW laser and the comb are obtained at each wavelength by blocking the appropriate arm and recording a new trace.

As expected, comb lines closest to the 1550-nm pump wavelength exhibit the highest degree of contrast (nearly 60~dB at 1319~nm).  However, even at the edge of the spectrum near 775~nm, more than 35~dB of contrast is obtained.  In all cases the spectra exhibit very clean modes without any visible inter-mode artifacts or sidebands.

\begin{figure}
\centering
\includegraphics[width=0.7\linewidth]{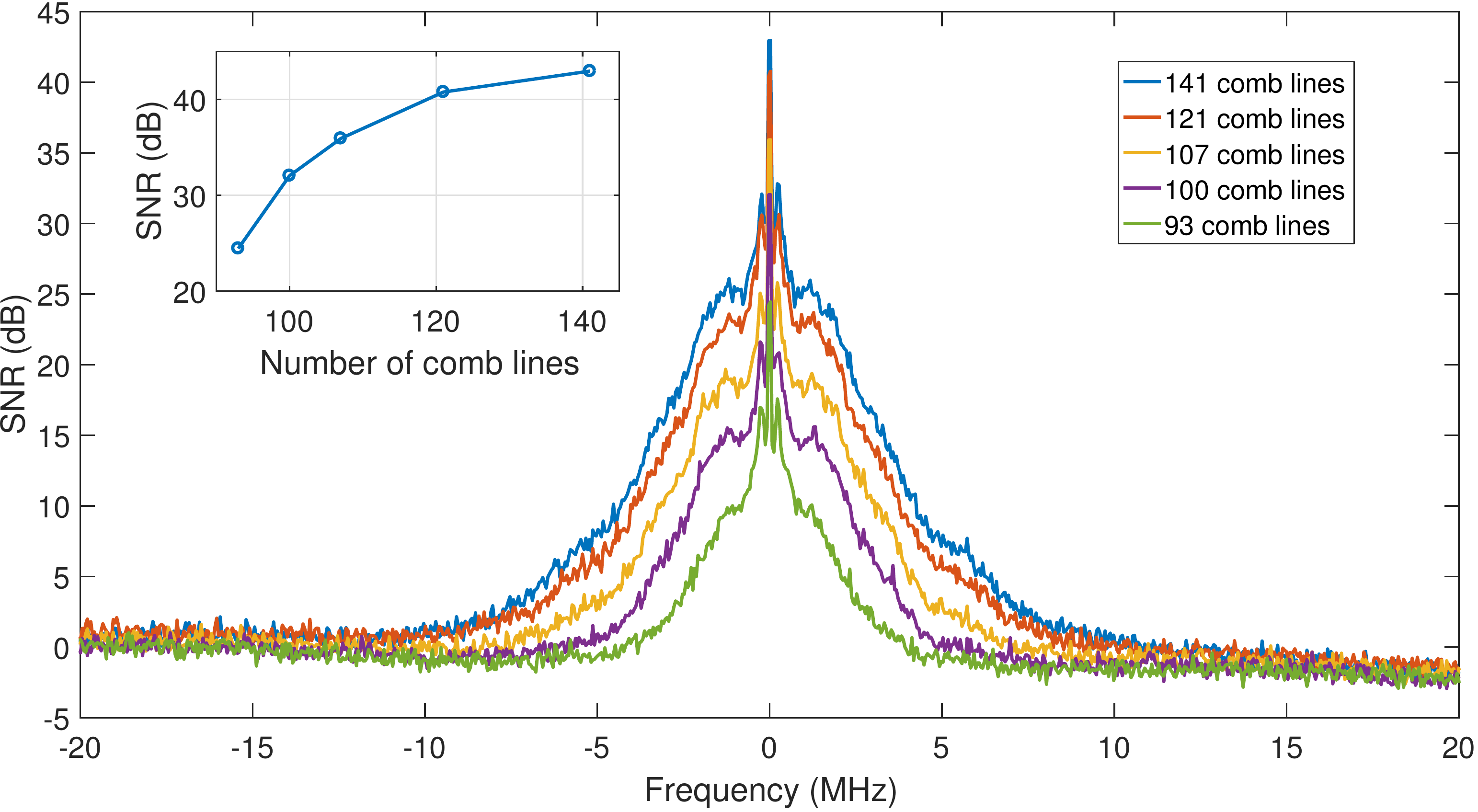}
\caption{Offset-frequency beat notes of the 10-GHz EOM comb as a function of the number of modes in the initial spectrum.  Narrower-bandwidth comb spectra suffer from reduced coherence after supercontinuum generation. The inset shows beat note SNR versus the number of initial comb lines.}
\label{fig:offsetvslines}
\end{figure}

As discussed in the main text, broad supercontinuum generation using narrow-bandwidth seed pulses poses significant challenges for low-noise performance.  In conventional all-anomalous-dispersion media, modulation instability can lead to noisy spectra and degradation of the comb coherence.  Similar effects are seen in our system, though the problem is strongly mitigated by the first-stage broadening in normal-dispersion HNLF.  To demonstrate this, Fig. \ref{fig:offsetvslines} shows the offset-frequency beat notes as a function of EOM-comb bandwidth. Each curve in the plot is obtained by symmetrically filtering the initial comb spectrum in increasing amounts using a spatial light modulator. At each step, the amplifier power is adjusted to maximize the achievable signal-to-noise ratio (SNR) of the beat.  The SNR increases monotonically with the number of comb modes, indicating it is advantageous to maximize the microwave drive power and the number of phase modulators for best performance.

\subsection*{Reliability and continuous operation}

\begin{figure}
\centering
\includegraphics[width=0.5\linewidth]{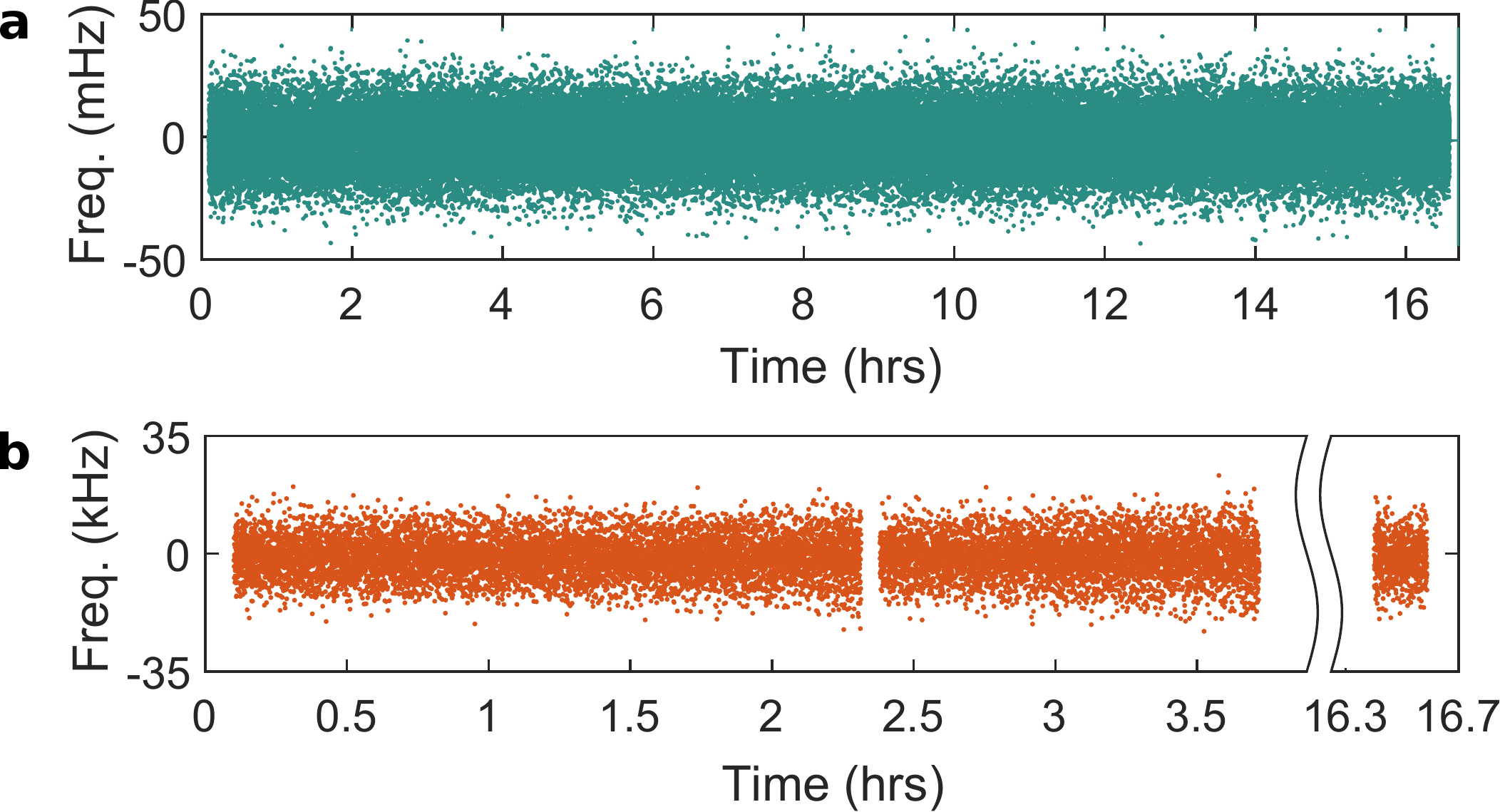}
\caption{Reliable overnight operation.  a) In-loop counter data showing continuous glitch-free operation for more than 16~hrs overnight (30~MHz set point subtracted). b) Out-of-loop absolute frequency verification during acquisition (offset by expected value of 30.6 MHz).  Time gaps in the data are due to the reference fiber comb losing phase lock.}
\label{fig:overnight}
\end{figure}

An important benefit to EOM-comb technology is its inherent reliability.  To support this claim, and the ability to generate absolute optical frequencies, we ran the system continuously overnight for more than 16~hrs.  In this configuration, a tunable external-cavity diode laser was used as the CW pump and was frequency-stabilized through feedback from the comb offset $f_0$.  The repetition rate, on the other hand, was generated by a low-noise microwave synthesizer, to which the DRO was phase locked.  Both comb parameters were derived from an SI-second-referenced hydrogen maser and thus provide absolute calibration of the comb modes.  Fig.~\ref{fig:overnight}a shows frequency counter data (gate time 1~s) of the in-loop comb offset frequency.  No glitches are observed during the entire data set.  

The absolute frequency of the comb is verified during this same acquisition by counting a heterodyne beat between the EOM-comb pump laser and an independent self-referenced fiber frequency comb (Fig.~\ref{fig:overnight}b).  Gaps in the record are due to the secondary comb exceeding the range of its stabilization actuators and losing phase lock.

\subsection*{Accuracy and stability measurement}
To assess the comb accuracy and stability, shown in Fig.~\ref{fig:accuracy} in the main text, the EOM comb repetition rate was compared against an auxiliary self-referenced comb. This secondary comb was a commercial fiber laser ($f_{\rm{rep,aux}} = 250$~MHz) and was phase locked to the same 1550-nm CW laser that serves as the pump for the EOM comb.  The repetition rate of the EOM comb was electronically mixed with the 40$^{\rm{th}}$ harmonic of the auxiliary comb's repetition rate at 10~GHz, resulting in a beat at $f_{\rm{diff}}$ that was counted with a frequency counter.  The exact value of $f_{\rm{diff}}$ can be calculated with knowledge of the comb mode numbers and the phase-lock frequency set-points, as shown below.

The EOM-comb pump-laser frequency $\nu_p$ is first expressed in terms of $f_0$ and $f_{\rm{rep}}$ for each comb:
\begin{eqnarray*}
\nu_p &=& f_{0} + N f_{\rm{rep}} \\
\nu_p &=& f_{0,\rm{aux}} + M f_{\rm{rep,aux}} - f_a
\end{eqnarray*}
where $N$ and $M$ are the mode numbers at the pump frequency for the EOM comb and auxiliary comb, respectively.  $f_a$ is the frequency-offset set point used to phase lock the auxiliary comb to the CW laser.

Rewriting the comb equations in terms of the repetition rates yields:
\begin{eqnarray*}
f_{\rm{rep}} &=& \frac{1}{N}(\nu_p - f_{0}) \\
f_{\rm{rep,aux}} &=& \frac{1}{M}(\nu_p - f_{0,\rm{aux}} + f_a).
\end{eqnarray*}

Subtracting $f_{\rm{rep}}$ from the 40$^{\rm{th}}$ harmonic of $f_{\rm{rep,aux}}$ yields an expression for $f_{\rm{diff}}$ in terms of the frequency set points, mode numbers, and pump frequency:
\begin{eqnarray*}
f_{\rm{diff}} &=& 40f_{\rm{rep,aux}} - f_{\rm{rep}} \\
		 &=& \left(\frac{40}{M} - \frac{1}{N}\right)\nu_p  + \frac{40}{M}(fa - f_{0,\rm{aux}}) + \frac{1}{N}f_{0}.
\end{eqnarray*}

In the experiment, $N = 19339$.  Thus, by tuning the auxiliary comb such that $M = 40\times N = 773\:560$, the $\nu_p$ term is entirely canceled.  Using the experimental values of $f_a = 40$~MHz, $f_{0,\rm{aux}} = 30$~MHz, and $f_0 = 3542$~MHz yields:

\begin{eqnarray*}
f_{\rm{diff}} &=& \frac{1}{N}(f_a-f_{0,\rm{aux}}+f_{0}) \\
&=& 183\:670.\:303\:531\:723\:450\ \rm{Hz}.
\end{eqnarray*}

\end{document}